\documentclass[aps,prl,twocolumn,a4paper,10pt,notitlepage,superscriptaddress,final,longbibliography]{revtex4-1}
\usepackage[english]{babel}
\usepackage{endnotes}
\usepackage{amssymb,amsmath,amsfonts}
\usepackage{textcomp}
\usepackage{graphicx,color}
\newcommand{\norm}[1]{\lvert#1\rvert}%

\newcommand{\New}[1]{\textcolor{green}{#1}}%

\renewcommand{\vec}[1]{\boldsymbol{\mathrm{#1}}}%

\newcommand{\DT}{D_{\mathrm{T}}}%
\newcommand{\DR}{D_{\mathrm{R}}}%
\newcommand{\uu}{\boldsymbol{\mathrm{\hat{u}}}}%

\begin{document}

\pagestyle{plain}

\title{Light-induced self-assembly of active rectification devices}

\author{Joakim Stenhammar}
\email{joakim.stenhammar@fkem1.lu.se}
\affiliation{Division of Physical Chemistry, Lund University, 221 00 Lund, Sweden}

\author{Raphael Wittkowski}
\affiliation{Institut f\"ur Theoretische Physik II, Heinrich-Heine-Universit\"at D\"usseldorf, D-40225 D\"usseldorf, Germany}

\author{Davide Marenduzzo}
\affiliation{SUPA, School of Physics and Astronomy, University of Edinburgh, Edinburgh EH9 3FD, United Kingdom}

\author{Michael E. Cates}
\affiliation{DAMTP, Centre for Mathematical Sciences, University of Cambridge, Cambridge CB3 0WA, United Kingdom}

\date{\today}

\begin{abstract}
Self-propelled colloidal objects, such as motile bacteria or synthetic microswimmers, have microscopically irreversible individual dynamics -- a feature they share with all living systems. The incoherent behavior of individual swimmers can then be harnessed (or ``rectified'') by microfluidic devices that create systematic motions impossible in equilibrium. Here we present a computational proof-of-concept study, showing that such active rectification devices might be created directly from an unstructured ``primordial soup'' of light-controlled motile particles, solely by using spatially modulated illumination to control their local propulsion speed. Alongside both microscopic irreversibility and speed modulation, our mechanism requires spatial symmetry breaking, such as a chevron light pattern, and strong interactions between particles, such as volume exclusion causing a collisional slow-down at high density. Taken together, we show how these four factors create a novel, many-body rectification mechanism. Our work suggests that standard spatial-light-modulator technology might allow the programmable, light-induced self-assembly of active rectification devices from an unstructured particle bath. \\[5mm]
\textbf{One-sentence summary:} Self-propelled particles that swim in response to light can self-assemble microfluidic rectification devices under non-uniform illumination.
\end{abstract}
\maketitle

\thispagestyle{plain}

\section{INTRODUCTION}
The last decade has seen a surge of interest in so-called ``active'' materials, whose constituent building blocks violate time-reversal symmetry on the microscopic scale by continuously converting fuel into motion \New{\cite{Marchetti-2013,Popkin2016}}. These building blocks can be natural, such as swimming bacteria or algae~\cite{Cates-2012}, or synthetic, such as self-phoretic Janus swimmers~\cite{Howse-2007,Brown-2014}. Apart from deepening our understanding of nonequilibrium systems, and in particular about how and when concepts from equilibrium thermodynamics can (and cannot) be applied to such systems~\cite{Cates-2015,Solon-2015}, a strong motivation behind active matter research is that activity can be harnessed to create devices and induce phenomena that would be impossible in thermodynamic equilibrium. Examples include flow of rotor particles round a circuit~\cite{Bartolo-2013}, steady rotation of a gear wheel in a bacterial bath~\cite{Angelani-2009,DiLeonardo-2010}, and pumping of bacteria between chambers by ``funnel gates''~\cite{Galajda-2007}. 

In this study, we show how irreversible motion in a particular class of active matter -- light-controlled self-propelled particles -- can be used for self-assembling active ``rectification devices'' and circuits directly from an unstructured active-particle suspension using only spatially controlled (but possibly incoherent) lighting. Such light-controlled motility has been experimentally demonstrated several times, both in synthetic systems \cite{Palacci-2013,Buttinoni-2013,KuemmeltHWBEVLB2013,Palacci-2014} and natural ones (strains of {Escherichia} coli bacteria whose motility requires illumination) \cite{Walter-2007}. The possibility of such light-controlled self-assembly is far from obvious, since light intensity is, for our purposes, a scalar rather than a vector field: it controls the speed of particles, but not their orientation; therefore, uniform illumination does not itself create directional motion. This differs fundamentally from magnetotaxis, for example, in which a uniform vector field directs motion by aligning the propulsion direction of particles \cite{Blakemore-1982}. Because of the ubiquity of spatial-light-modulator (SLM) technology \cite{Grier-2003}, the ability to create active rectification devices using light-induced self-assembly could have many advantages over conventional microfluidic approaches to rectification \cite{Angelani-2009,DiLeonardo-2010,Galajda-2007,Wan-2008,Koumakis-2013,Koumakis-2014}. In particular, it would allow real-time device reprogramming rather than mechanical interchange of fixed microfluidic elements. Our proposed strategy hinges on the tendency of motile particles to accumulate in regions where they move slowly \cite{Schnitzer-1993,Tailleur-2008,Cates-2013}. This  allows one to use light to sculpt regions of high and low particle density. So long as the particles interact repulsively, the dense regions might be used to create, in effect, microfluidic obstacles of a type previously used for rectification \cite{Galajda-2007}. We first develop this appealingly simple avenue before turning to other approaches that prove more efficient. 

The slowing-induced accumulation of motile particles is much like the crowding of pedestrians who slow down in front of a shop window on a busy street \cite{Cates-2015}. Within a wide class of models \cite{Cates-2013}, it is quantified as $\rho(\mathbf{r}) \propto 1/v(\mathbf{r})$, where $\rho(\vec{r})$ is the steady-state mean particle density at position $\vec{r}$ and $v(\vec{r})$ is the local swim speed \cite{Schnitzer-1993}. This result has no counterpart for nonmotile (e.g., Brownian) particles in thermal equilibrium, whose steady-state density is governed solely by the Boltzmann distribution, and whose speed statistics remain completely independent of their position. It applies rigorously for noninteracting particles \cite{Schnitzer-1993,Cates-2013} and generalizes as $\rho(\vec{r}) \propto 1/\overline v(\vec{r},\rho)$ to cases where interactions cause the average local swim speed $\overline{v}(\vec{r},\rho)$ to differ from the local value for an isolated particle, $v(\mathbf{r})$, which we assume to be illumination-controlled \cite{Tailleur-2008,Cates-2013}.

\section{RESULTS AND DISCUSSION}
\subsection{Light-controlled accumulation of active particles}
\begin{figure}[ht] 
\begin{center}
\includegraphics[width=\linewidth]{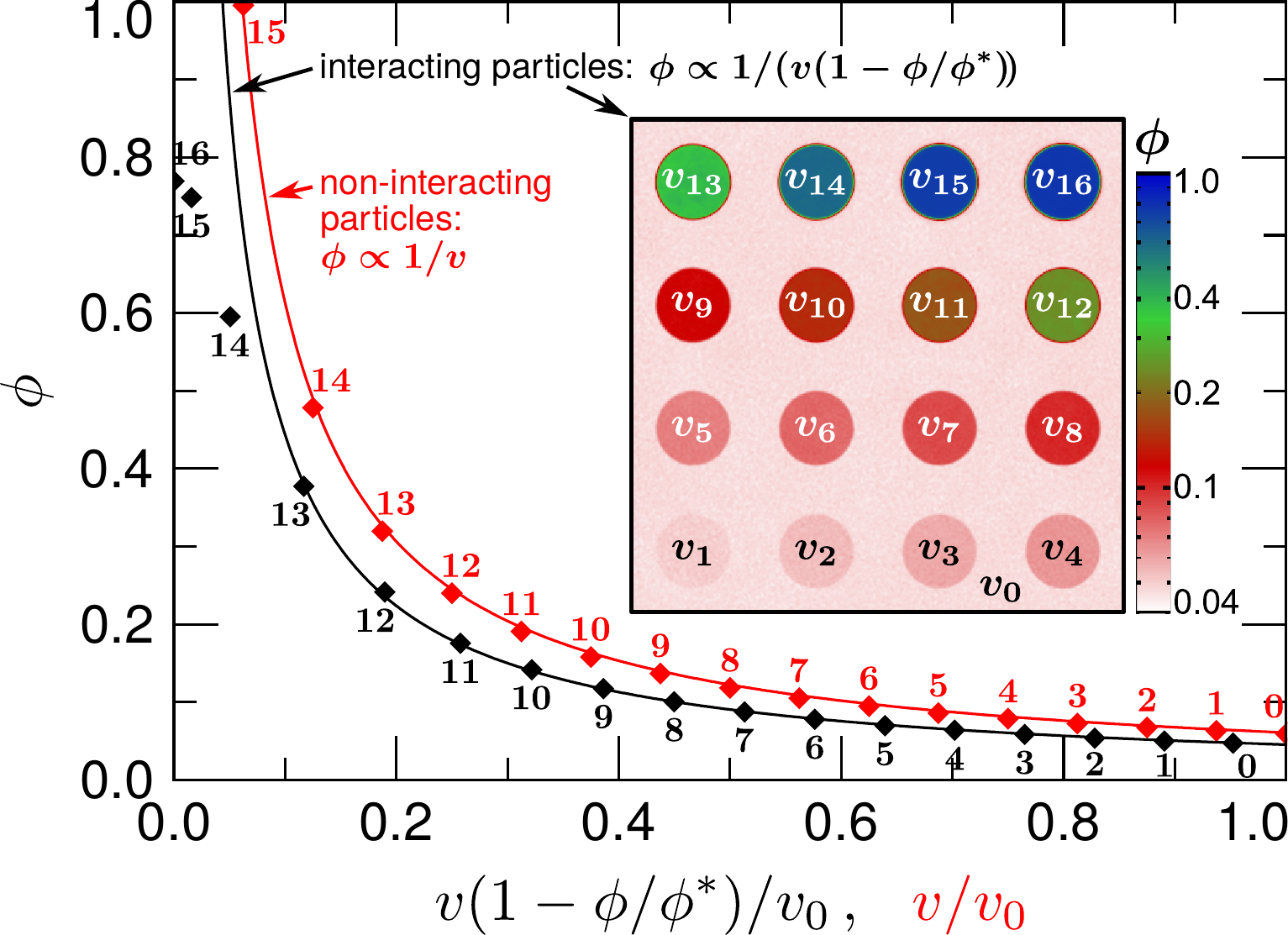}
\caption{\label{lattice}Time-averaged local volume fraction $\phi(\vec{r})$ of interacting and noninteracting particles in a system with spatially patterned one-particle swim speed $v(\mathbf{r})$, where $v=v_i=v_0(1-i/16)$ in the $i$th well and $v=v_0$ between the wells. The solid lines show our predictions with a prefactor determined by least-squares fits, where we neglected the three highest density ($\phi>0.5$) data points for the interacting particles. Inset: Spatially resolved plot of $\phi(\vec{r})$ for the interacting particles; note the logarithmic scale on the color bar.}
\end{center}
\end{figure}

For our proof-of-concept simulations we deploy a well-studied model of ``active Brownian particles'' in two spatial dimensions. Each particle swims with a constant speed $v$ along a particle-fixed axis that rotates by autonomous angular diffusion; this motion is influenced by interparticle collisions (see Methods). It is known that at fixed, spatially uniform $v$, collisions cause a linear dependence of the average swim speed on density: $\overline{v}(\phi) = v(1-\phi/\phi^{*})$ with $\phi = \rho \pi \sigma^2/4$ the volume fraction of particles in two dimensions, $\sigma$ the particle diameter, and $\phi^{*} \approx 0.93$ a near-close-packed volume fraction that depends on interaction details \cite{Stenhammar-2013,Stenhammar-2014}. Hence, in this system one might expect that if a spatial speed-modulation pattern $v(\mathbf{r})$ is imposed, $\phi(\mathbf{r})$ should everywhere obey $\phi \propto 1/(v(1-\phi/\phi^*))$. This prediction is tested in Fig.\ \ref{lattice} for a modulation pattern comprising a lattice of $16$ circular traps, each with a different uniform nominal speed $v_i = v_0(1-i/16)$, $i=1,\dotsc,16$, in its interior and a larger speed $v_0$ between the traps. (We use periodic boundary conditions.) The agreement is excellent, except at the highest densities where, with the chosen interactions (see Methods), particles show significant overlap, invalidating the assumed linear dependence of $\overline{v}$ on $\phi$ \cite{Stenhammar-2013}.

\subsection{Self-rectification by funnel gates}
The above results demonstrate that for motile particles with repulsions, just as for noninteracting ones, the density pattern $\rho(\mathbf{r})$ can be rather precisely manipulated by spatial modulation of the one-particle swim speed $v(\mathbf{r})$. But Fig.\ \ref{lattice} does not demonstrate rectification: that would require either a continuous steady-state particle current, or, if the boundary conditions prevent such a current, two areas of the same $v$ to acquire different densities. Only then can one claim to have converted incoherent particle dynamics, which by itself is responsible for the $\rho \propto 1/\overline v$ behavior, into a systematic pumping effect.

\begin{figure}[ht] 
\begin{center}
\includegraphics[width=\linewidth]{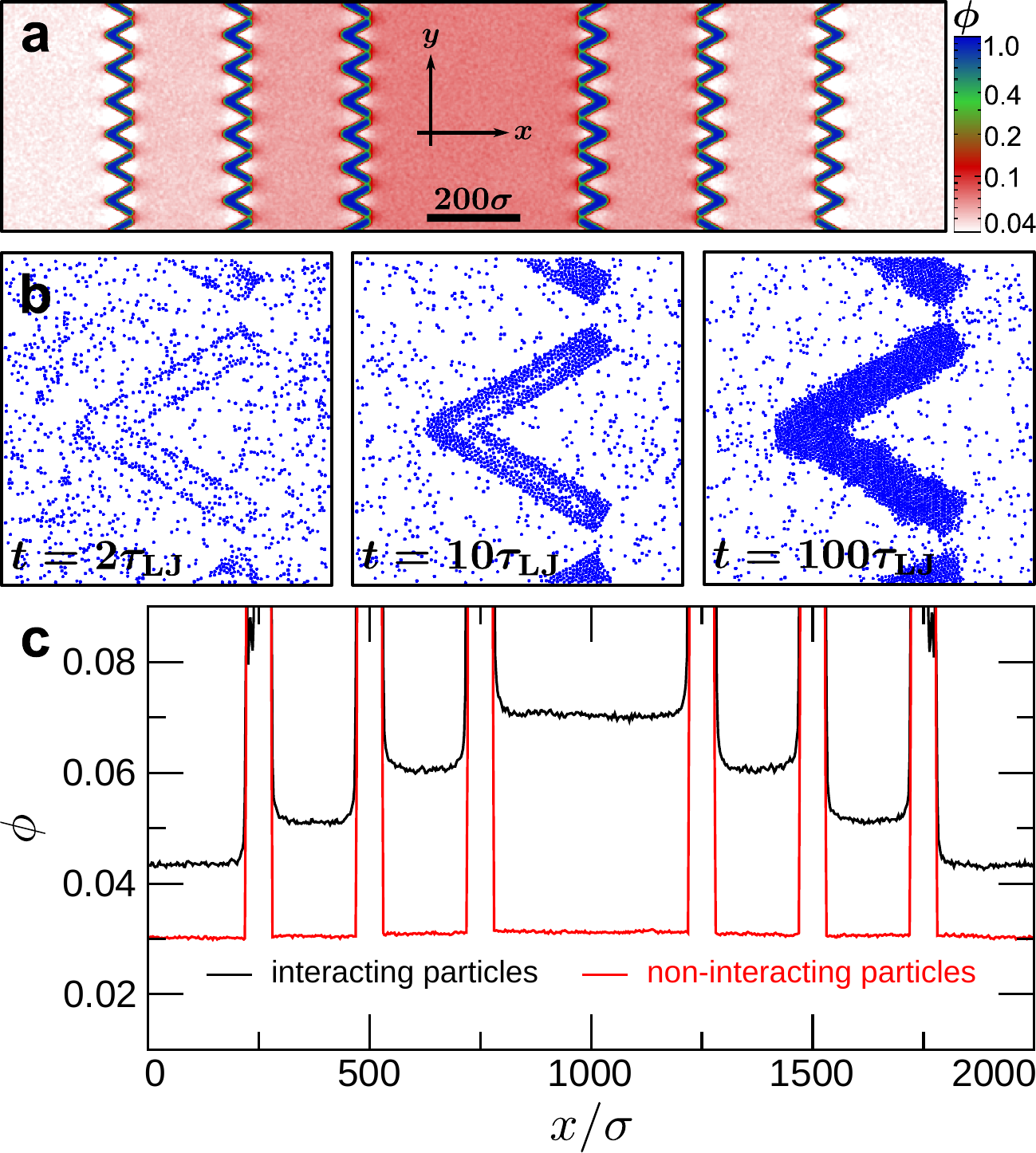}
\caption{\label{funnel}(a) Time-averaged particle packing fraction $\phi(\vec{r})$ at steady state in a system of self-assembled funnels. Note the logarithmic scale on the color bar. (b) Magnification of one of the funnels, showing the self-assembly process at three different times starting from a homogeneous particle distribution; $\tau_{\mathrm{LJ}}$ is the Lennard-Jones time (see Methods). (c) Black curve: The packing fraction $\phi$ from (a), but now as a function of $x$ and averaged over $y$; the pumping effect (rectification) is obvious. Red curve: For noninteracting particles, rectification is not possible.}
\end{center}
\end{figure}

A simple strategy for achieving such rectification is to create a light-guided density pattern that mimics microfluidic devices. Such devices generally operate at the one-particle level: a series of obstacles is created so that each particle is geometrically biased toward a nonzero average motion. However, when the obstacles are made of other particles this becomes an inherently many-body mechanism that computer simulations can illuminate. In an elegant experimental demonstration of one-particle rectification, Galajda \emph{et al.\ }\cite{Galajda-2007} established that a wall of microfabricated chevron-shaped obstacles or ``funnels'' will rectify a suspension of swimming \emph{Esch\-e\-ri\-chia coli} bacteria. Asymmetric pumping across such a wall between two chambers of the same $v$ can yield more than a factor-two density difference. The effect of additional layers of walls is multiplicative on the density ratio  \cite{Galajda-2007}.

Inspired by these experiments, we now simulate a set of chevron-shaped ``dark'' regions, in which the nominal swim speed $v(\mathbf{r})$ is set to zero. As can be seen in Fig.\ \ref{funnel} and Movie S1, this pattern rapidly guides the self-assembly of a set of funnel-like obstructions which in turn cause rectification. The mechanism is almost the same as in Refs.\ \cite{Galajda-2007,Wan-2008}: the trajectories of motile particles will tend to trace the funnel wall, making the probabilities of passing through the funnel from either side unequal. However, because our funnels assemble reversibly from the motile particles themselves, such rectification requires not only time-irreversible trajectories (wall-tracing) and broken reflection symmetry of the modulation pattern \cite{Julicher-1997,Wan-2008,Tailleur-2009} but also repulsive interparticle interactions (Fig.\ \ref{funnel}). Unlike conventional microfabricated structures, ours could in principle be dismantled and re-assembled into a completely different structure, simply by reprogramming the light intensity pattern to give a different $v(\mathbf{r})$. Crucially, the mechanism requires no external or interparticle forces that align the swimming directions of the particles. 

As a first proof of principle, these are striking results. However, the efficiency of our device (measured as the ratio between the densities in the inner and the outer compartments) is modest: the density ratio induced by three pairs of funnel walls is about 1.6 (see Fig.\ \ref{funnel}c), compared to about 2.8 seen in \emph{Escherichia coli} experiments using a single microfabricated funnel wall \cite{Galajda-2007} and in simulations of rod-shaped particles in a similar geometry \cite{Wan-2008}. As shown in the Supplementary Material, this is largely because our self-assembled many-particle funnels cannot prevent particles from ``leaking'' backwards against the pumping flux, limiting their pumping efficiency.

\subsection{Self-rectification by one-dimensional speed profiles}
\begin{figure}[ht]
\begin{center}
\includegraphics[width=\linewidth]{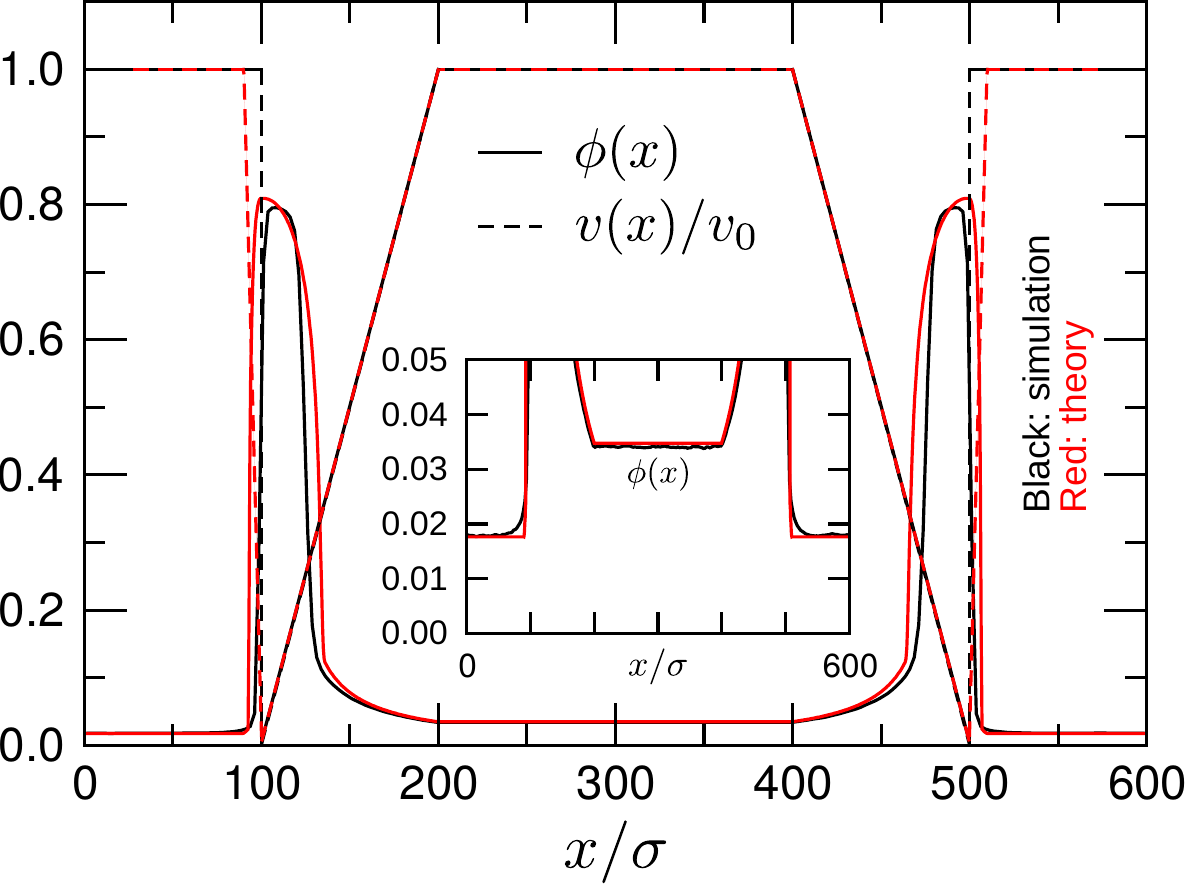}
\caption{\label{density_ratchet} Steady-state packing fraction $\phi(x)$ (solid curves) in a system with a modulated speed $v(x)$ composed of two opposing saw-tooth profiles (dashed curves), obtained from particle-resolved simulations and our continuum model. For numerical reasons, the discontinuities of $v(x)$ are smeared out over $10\sigma$ in the continuum calculations. The inset shows $\phi(x)$ with a magnified $y$ scale, clearly demonstrating the rectification.}
\end{center}
\end{figure}

A contrasting single-particle rectification strategy deploys microfabricated wedge-shaped solid barriers with one steep and one shallow side. Such barriers have been shown experimentally \cite{Koumakis-2013,Koumakis-2014} to make swimming bacteria preferentially transport passive tracer colloids from the shallow side and deposit them on the steep side of the barrier, leading to a significant steady-state accumulation. Loosely inspired by this, we abandoned the chevron design, and imposed a simpler, one-dimensional sawtooth-shaped speed profile $v(x)$ shown in Fig.\ \ref{density_ratchet}. This leads to the self-assembly of an asymmetric barrier with a distinctive density profile $\rho(x)$. 

Remarkably, this extremely simple device shows a density ratio of almost $2$ and thus a strong rectification (see inset of Fig.\ \ref{density_ratchet}). Note that particles must pass through, but not over or around, the barrier so that the ``leakage'' encountered earlier is being exploited here. To study this rectification, we have extended a continuum model, involving local density and orientation fields coupled in a standard manner \cite{Farrell-2012}, to the case where the swim speed depends on both position and density (see Methods). The rectification is found to arise from a polarization (i.e., alignment of particle orientations) on the dense (steep) side of the barrier; here the average swimming direction of the particles is oriented into the barrier (see the Supplementary Material for details). This polarization is a straightforward consequence of the irreversible particle motion, which makes particles that swim towards a solid object such as a wall become ``trapped'' there until their swimming direction has rotationally diffused enough to let them escape, creating an average particle orientation pointing into the object. The polarization in turn creates a mechanical pressure that drives a particle flux from the steep side of the barrier towards the shallow side. In more mathematical terms, rectification rests on the presence of a density-gradient-driven pumping effect that is nonzero \emph{only} in a small region close to the barrier. Flux balance then requires a change in the uniform steady-state particle densities in the bulk regions far to the left and to the right of the barrier, where density gradients, and thus the pumping effect, approach zero. In contrast to equilibrium systems, here local gradient effects cause rectification across an interface between bulk regions, rather than merely shifting the interfacial tension \cite{Wittkowski-2014}.

\subsection{Self-assembled active circuits}
\begin{figure}[ht] 
\begin{center}
\includegraphics[width=\linewidth]{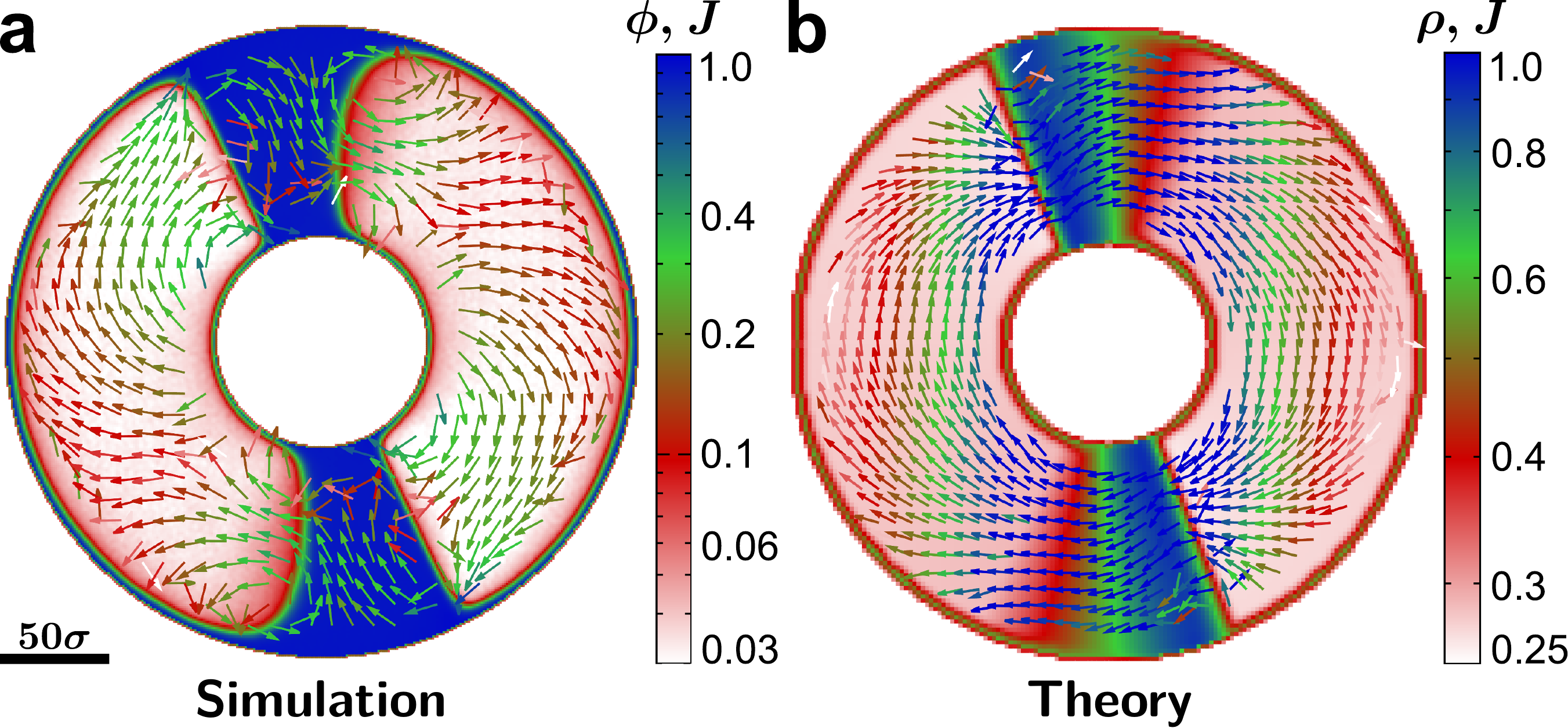}
\caption{\label{circuit}(a) Particle-resolved simulations of an annular circuit self-assembled through a prescribed speed profile $v(\vec{r})$ with two similar saw teeth along the annulus and a very low speed outside of the annulus. The resulting dense particle packing outside of the annulus (not shown for clarity) confines the motile particles to the circuit. The color map shows the time-averaged steady-state local packing fraction $\phi(\vec{r})$ and vectors denote the rescaled local particle current $\vec{J}(\vec{r})$, with their colors indicating the absolute current density $J(\vec{r})=\norm{\vec{J}(\vec{r})}$. The logarithmic color bar applies both to the density plot and to the vector field plot.
(b) Corresponding plot of the rescaled particle density $\rho(\vec{r})\propto\phi(\vec{r})$ and particle current $\vec{J}(\vec{r})$ obtained from our continuum model.}
\end{center}
\end{figure}

By using aligned (instead of opposing) barriers and applying suitable boundary conditions, either of the above devices can be reconfigured to create an ``active circuit'', where motile particles flow endlessly around a closed loop at steady state. In Ref.\ \cite{Bartolo-2013} macroscopic circulation was achieved with field-driven autorotors with hydrodynamic alignment interactions. In Fig.\ \ref{circuit} we give computational proof-of-principle for a simpler design that employs the speed modulation from Fig.\ \ref{density_ratchet}, but now with two aligned saw teeth. In this example, colloids are confined to an annulus by a surrounding dark region; thus, both the ``circuit'' itself and the rectifying elements arise solely by light-guided self-assembly. 

\section{CONCLUSION}
In conclusion, we have given proof-of-principle evidence, using computer simulations and theoretical considerations, that self-assembled active rectification devices for motile particles can be created solely through spatial modulation of the local propulsion speed $v(\mathbf{r})$. This space-dependent propulsion speed can in turn be realized by a space-dependent illumination \cite{Palacci-2013,Palacci-2014,Buttinoni-2013,Walter-2007} and hence rendered programmable through use of SLMs. A key element of the light-guided assembly strategy presented here is that the illumination supplies the particles only with \emph{scalar} information, namely the instruction to swim at a certain speed. In some experimental systems the light elegantly also supplies the energy source for swimming \cite{Buttinoni-2013}, but this is not always the case \cite{Palacci-2013}, and is not an essential ingredient in our proposed strategy. Crucially, in our mechanism the optical field does not exert directional forces directly on the particles. This differs fundamentally from, and requires much lower light intensities than, the manipulation of particles by optical tweezers; as an example, the light-propelled particles of Ref.\ \cite{Buttinoni-2013} require light intensities of $\leq 5$ $\mu$W$/\mu$m$^{2}$, compared to intensities of $10^{2}-10^{4}$ $\mu$W$/\mu$m$^{2}$ typically used in optical tweezers \cite{Jonas-2008,LechnerPMDK2015}. Our mechanism can be easily applied to large-scale systems containing many particles, circumventing several disadvantages of traditional micro-manipulation techniques \cite{Jonas-2008}, such as significant heating-up of the system and the need for complicated optical equipment. Indeed, although use of SLMs would certainly be appealing, there is no requirement that the light field even be coherent. 

Our work points towards a new technological opportunity in which light-guided self-assembly is not only used to make crystals and other aggregates \cite{Palacci-2013,Palacci-2014,Buttinoni-2013} but also to create programmable, active rectification devices from an initially structureless ``primordial soup'' of self-propelled particles. The capacity to do this stems from an interplay between microscopically irreversible particle motion, spatially asymmetric light modulation, and interparticle repulsions. These findings may also be relevant for controlling mass transport in biological systems, steering pattern formation in future applications of active particles in materials science, and the wider deployment of active particles in colloidal self-assembly.

\section{METHODS}
\subsection{Particle-based model} 
Our particle-resolved computational studies use a simple model of swimming microorganisms (like motile bacteria and protozoa) and artificial microswimmers, known as ``active Brownian particles''. Such particles exhibit persistent propulsion in a particle-fixed direction which rotates through thermal rotational diffusion. This model has been shown to capture many essential features of active matter systems such as bacterial suspensions, while still allowing for both the derivation of exact results from first principles and computational studies of systems containing millions of particles. The active colloids are modeled as spherical particles with diameter $\sigma$ moving on a two-dimensional substrate. They interact repulsively through a truncated and shifted Lennard-Jones potential $U(r) = 4\varepsilon ((\sigma/r)^{12} - (\sigma/r)^{6}) + \varepsilon$ with the center-to-center particle distance $r$ and an upper cut-off at $r = 2^{1/6}\sigma$, beyond which $U = 0$. $\varepsilon$ determines the interaction strength and $\beta = 1/(k_{\mathrm{B}}T)$ is the inverse thermal energy. The motion of these particles is described by the overdamped Langevin equations
\begin{align}
\dot{\vec{r}}_i &= v(\vec{r}_i)\uu(\theta_i) + \beta \DT \vec{F}_i(\{\vec{r}_j\}) + \sqrt{2\DT } \:\! \boldsymbol{\Lambda}_{\mathrm{T}} \;, \label{Langevin_t} \\
\dot{\theta}_i &= \sqrt{2\DR } \:\! \Lambda_{\mathrm{R}} \;, \label{Langevin_r}
\end{align}%
where $\mathbf{r}_{i}(t)$ is the position and $\theta_{i}(t)$ is the orientation of the $i$th particle at time $t$. In our particle-resolved Brownian dynamics simulations we solved these Langevin equations numerically using the LAMMPS molecular dynamics package \cite{Plimpton-1995}. Here, $v(\vec{r})$ is the nominal swim speed of an active particle at position $\vec{r}$, $\uu(\theta)=(\cos(\theta),\sin(\theta))^{\mathrm{T}}$ is an orientational unit vector that denotes the propulsion direction of a particle with orientation $\theta$, and $\vec{F}_i(\{\vec{r}_j\})$ is the total conservative force on particle $i$, which results from the interactions with other particles and thus depends on $U(r)$. $\DT $ and $\DR  = 3\DT /\sigma^{2}$ denote the constant translational and rotational diffusion coefficients of the particles, respectively, and $\boldsymbol{\Lambda}_{\mathrm{T}}(t)$ and $\Lambda_{\mathrm{R}}(t)$ are zero-mean, unit-variance Gaussian white noise terms. The natural time unit in this model is the Lennard-Jones time $\tau_{\mathrm{LJ}}=\sigma^{2}/(\varepsilon\beta \DT )$. Further technical details about the particle-resolved simulations can be found in the Supplementary Material.

\subsection{Continuum model} 
In our continuum simulations we solved the following dynamic equations for the local particle density $\rho(\vec{r},t)$ and the local polarization $\vec{P}(\vec{r},t)$ numerically:
\begin{align}
\dot{\rho} = & - \nabla\!\cdot\!\left(\overline{v} \mathbf{W}\right) + \DT  \nabla^2 \rho \;, \label{continuum_rho}\\
\dot{\vec{W}} = & - \frac{1}{2}\nabla (\overline{v} \rho) -\gamma_1 \mathbf{W} - \gamma_2 W^2 \mathbf{W}  + k \nabla^2 \mathbf{W} \label{continuum_W}
\\ \nonumber & - w_1 (\mathbf{W}\!\cdot\!\nabla) \mathbf{W} + w_2 \nabla (W^2) \;. 
\end{align}
Here, $\vec{W}=\rho\vec{P}$ is a weighted local polarization with modulus $W=\norm{\vec{W}}$ and $\gamma_1$, $\gamma_2$, $k$, $w_1$, and $w_2$ are parameters. 
This continuum model can be obtained by an appropriate adaptation of the derivation of a mean-field theory for a suspension of active particles in Ref.\ \cite{Farrell-2012}. 
The average swim speed $\overline{v}(\mathbf{r},\rho)$ of an active particle at position $\vec{r}$ in an active suspension of density $\rho$ is known from particle-resolved simulations \cite{Stenhammar-2013,Stenhammar-2014} to be linearly decreasing with $\rho$. Nevertheless, in our numerical calculations based on the continuum model we used the expression $\overline{v}(\mathbf{r},\rho) = v(\mathbf{r}) \exp(-\alpha \rho(\vec{r},t))$, where the parameter $\alpha$ determines how quickly the average local swim speed decreases with density. This choice increases numerical stability without changing the qualitative conclusions drawn here. 

The parameters $\gamma_1$ and $\gamma_2$ are positive and ensure that, in steady state and in absence of any density or velocity gradients, $\mathbf{W}$ vanishes everywhere.
$k$ is an effective diffusion coefficient for $\mathbf{W}$ and ensures continuity of this field. The term proportional to $w_1$ encodes self-advection, whereas the term proportional to $w_2$ can be viewed as an ``active pressure'' term, as it can be written as $-\nabla p$ with $p=-w_2 W^2$. 
Both these terms generically arise when coarse graining models of self-propelled particles \cite{Vicsek-1995,Toner-1995,Toner-1998,Farrell-2012}; the ensuing analysis suggests $w_2 > 0$ \cite{Farrell-2012}. 
Furthermore, if one considers $\mathbf{W}$ as a fast variable and performs a quasi-stationary approximation \cite{Cates-2013}, the $w_2$ term leads to a term proportional to $-w_2 \nabla^2 ((\nabla(\overline{v}\rho))^2)$ in Eq.\ \eqref{continuum_rho}; 
its form is similar to that of the leading-order active contribution in the minimal field theory for active-particle phase separation (``Active Model B'') \cite{Wittkowski-2014}. Further technical details about the numerical solution of the continuum model can be found in the Supplementary Material. 

\section{SUPPLEMENTARY MATERIAL} 
Supplementary material for this article is available at [\textit{URL to be inserted by the editor}]. \\
\textbf{Movie S1:} Movie showing the self-assembly process of a single chevron-shaped obstacle. \\
\textbf{Fig.\ S1:} Steady-state density profiles for funnel-wall systems with the funnel particles either mobile or fixed. \\ 
\textbf{Fig.\ S2:} Local polarization around a sawtooth-shaped speed profile, obtained from particle-resolved simulations and the continuum model. \\
\textbf{Fig.\ S3:} Polarization profiles obtained from the continuum model with $w_2 = 0$ and $w_2 > 0$. \\
\textbf{Fig.\ S4:} Propulsion speed profiles used in simulations of ``active circuits''.

\bibliographystyle{ScienceAdvances}

\vskip3mm

\textbf{Acknowledgments:} We thank Jochen Arlt, Clemens Bechinger, Aidan Brown, Nick Koumakis, Hartmut L\"owen, Vincent Martinez, and Wilson Poon for helpful discussions. 
\textbf{Funding:} This work was funded in part by EPSRC Grant EP/J007404. J.S.\ is financially supported by the Swedish Research Council (Vetenskapsr\aa det, Grant No.\ 350-2012-274), R.W.\ acknowledges financial support through a Return Fellowship (Grant No.\ WI 4170/2-1) from the Deutsche Forschungsgemeinschaft (DFG), and M.E.C.\ holds a Royal Society Research Professorship. 
\textbf{Author contributions:} All authors designed and performed the research and wrote the manuscript. 
\textbf{Competing interests:} The authors declare that they have no competing interests. 
\textbf{Data availability:} All data needed to evaluate the conclusions in the paper are present in the paper and/or the Supplementary Material. Additional data will be made available by J.S. upon request.

\end{document}